\documentclass{cernyrep}

\usepackage{color}
\newcommand{\aq}[1]{}
\newcommand{\tq}[1]{}
\newcommand{\ced}[1]{#1}

\usepackage{hyperref}

\begin{document}

\title{A Taste of Cosmology}

\author{L.~Verde\thanks{liciaverde@icc.ub.edu}}

\institute{ICREA and ICC,  Barcelona, Spain}

\maketitle

\begin{abstract}
This is the summary of two  lectures that aim to give an overview of cosmology.   I will not try to be  too rigorous in derivations, nor to give a full historical overview. The idea is to provide a ``taste'' of cosmology and some of  the interesting topics it covers.  The standard cosmological model is presented and  I highlight the successes of cosmology over the past decade or so. Keys to the development of the standard cosmological model are observations of the cosmic microwave background and of large-scale structure, which are introduced. Inflation and dark energy and the outlook for the future are also discussed.  Slides from the lectures are  available from the school website:
\url{physicschool.web.cern.ch/PhysicSchool/CLASHEP/CLASHEP2011/}.\tq{Do you wish to include the author's email address footnote and this link to the website?}
\end{abstract}

\section{Cosmology: Why should you care?}

Cosmology aims at the study of the Universe as a whole, but,  in particular, at getting an understanding of the Universe's origin, structure, composition, evolution and fate.
In general, and for many applications,  for cosmologists   galaxies can be treated as ``points''. This gives an idea of the scales involved.  Suitable units to measure scales must be introduced; cosmological distances are measured in megaparsecs, 1~Mpc = $3.86\times 10^{22}$~m; masses are often expressed in solar masses, $1~\mathrm{M}_{\odot}=1.99\times 10^{30}$~kg.

Cosmology in the past 15 years has made the transition to precision science and has moved from a data-starved science to a data-driven science. This development has been driven by new technology, which has made it possible to  survey  the Universe farther than before. As a result  of the ``avalanche of data'' of the past 20 years or so, cosmology now has a standard model. The ``standard cosmological model''  or the ``standard model of cosmology'' only needs a few parameters to describe the origin, composition and evolution of the Universe on large scales. However, there is a great difference between modelling and understanding. As we will see, the standard cosmological model leaves many deep open questions, for example: What is dark matter?  Do we understand gravity on very large scales? Is there energy associated with the vacuum? These questions show that there are deep connections between cosmology and fundamental physics, and that  cosmological observations  can be crucial to understanding open questions in physics.

Testing fundamental physics by looking up at the sky is not new. The interplay between astrophysics and fundamental physics has already produced spectacular findings (as an example, think about the solar neutrino problem).  Cosmological observations  are key to  testing fundamental physics in four areas, in particular: dark matter, neutrinos, inflation and dark energy.

\subsection{Assumptions}

There are some fundamental assumptions at the basis of  cosmology.
 \begin{enumerate}
\item Physics as we know it can describe the Universe as a whole, i.e.\ it is the same in all corners of the Universe.
\item On the largest scales -- those mostly of interest for cosmology -- the driving force is gravity, and we can neglect complicated effects of baryonic physics.
\item Gravity is described by general relativity (GR) -- although for many applications the Newtonian limit works fine.
\item The Universe on large scales is homogeneous and isotropic.  This is called the cosmological principle. We know that on small scales there are inhomogeneities and anisotropies -- think about galaxies, stars, the Solar System,  us \dots. But  on very large scales, one assumes that these inhomogeneities get smoothed, averaged out.
\item Therefore, the metric is  Friedmann--Lema\^{\i}tre--Robinson--Walker (FLRW). Locally, it cannot be, of course! Think about the metric next to a black hole, for example. This statement says that on very large scales the global  metric is FLRW.
\end{enumerate}
\noindent
Note that the puzzling aspects and the open questions  inherent to the standard cosmological model are leading many researchers to challenge these basic assumptions. While most people would feel uncomfortable touching the first two assumptions, an active area of research is to explore the observational consequences of dropping the other assumptions.

\subsection{Parameters of the model}

The parameters of the standard model for cosmology can be divided into three types: (i)~parameters describing the  smooth Universe, i.e.\ parameters that govern the global geometry of space-time and parameters that govern the expansion rate;  (ii)~parameters that describe the inhomogeneous Universe, i.e.\ that characterize the inhomogeneities; and (iii)~parameters that parametrize our ignorance -- for complicated, nonlinear physical processes that we cannot  properly describe and model, one  resorts to hiding all those and describing the effects with extra parameters.

Of course, it is important  to consider models more complex than the standard cosmological model by introducing ``extra parameters''.  Finding no evidence for these extra parameters would further support the standard model.

If the metric is FLRW, then there are only three possible  (global) geometries: closed, flat or open. The parameter that gives the global geometry is $\Omega=\rho /\rho_{\rm crit}$, where $\rho$ denotes the density. The critical density ($\rho_{\rm crit}$) is such that the geometry is spatial flat. $\Omega>1$ indicates a Universe with positive curvature and $\Omega<1$ one with negative curvature. $\Omega$ can be decomposed into several components, given by the composition of the  Universe.
For example, everything that shines (e.g.\ stars) is composed by baryonic matter. There is also, of course, dark baryonic matter  such as gas, dust, etc.

However, there is more than meets the eye.  The mass of galaxies, for example, can be computed from galaxy rotation curves (angular velocity plotted against distance from the centre).  The rotation curve of  a spiral galaxy can be measured by tracing the gas through specific  atomic or molecular transitions, usually  in the radio part of the electromagnetic spectrum. The rotation curves of galaxies remain flat well beyond the extent of the optical light, as first shown by Vera Rubin in the 1970s. In the Solar System, where most of the mass is concentrated in the Sun, the inner planets have higher circular velocity than the outer planets. A flat rotation curve  indicates the existence of an unseen  dark  component called a {\it halo} that extends well beyond the luminous part of a galaxy. It is useful to introduce a  concept called ``mass-to-light ratio'' to quantify the unseen (dark) matter component. Our Sun has by definition a mass-to-light ratio of unity.  Galaxies have a mass-to-light ratio of order~10.

Galaxies tend not to be isolated, but to cluster in {\it galaxy clusters} (gravitationally bound groups of hundreds of galaxies). In clusters of galaxies, the depth of the potential well (and thus the total mass) can be estimated from the velocity dispersion of the galaxies (as Zwicky did in the 1930s) or from the temperature of the intra-cluster gas, which is so hot that it emits X-rays.  This  gives a mass-to-light ratio for clusters of order~100, indicating  even more dark matter.  More spectacularly, and on even larger scales, the dark matter can be seen through its gravitational lensing: the deep potential well of the dark matter in clusters of galaxies bends the light path from background sources, distorting them into in arcs  and  making multiple images. From  gravitational lensing observations, the density profile of the dark matter in clusters (and in galaxies) has been mapped out. The dark matter ``halo'' extends well beyond the stars/galaxy component and  has a steep density profile  towards the centre.

More recently, new direct evidence has emerged on dark matter and its nature~\cite{refbullet}. The so-called ``bullet cluster'' shows the aftermath of an impact where a small cluster (the bullet) hit and passed through a larger cluster (the target). Combined observations of the optical (stars), X-rays (gas) and gravitational lensing (all, but mostly dark matter) show that the gas (baryonic matter, which is collisional), feeling drag due to electromagnetic interactions, lags behind the collisionless  component (dark matter and  stars/galaxies -- which on these large scales for all purposes act as a collisionless component).

Dark matter should be not only collisionless  but also cold, i.e.\ have small velocity.  Dark matter must be cold to give steep halo density profiles (see e.g.~\cite{NFW}) and to describe the clustering properties of large-scale structure (see section \ref{sec:lss}).

 Let us move on to consider the expansion of the Universe. Hubble in the 1920s realized that there was a  linear relation between the distance of galaxies and the wavelength shift of their spectra (mostly to longer wavelengths,  thus called redshift). By interpreting the galaxy redshift as recession velocity,  we obtain the Hubble law $cz=v=H_0d$, where $d$ denotes the distance, $z$ the redshift and $H_0$ the Hubble constant.  Owing to the finite speed of light, redshift is therefore an indicator not only of distance but also of time: light from an object  at ``high'' $z$ comes not only from  far away  but also from the past.

 Hubble showed that the Universe is not static but expanding (something that today we take for granted). As a consequence, the Universe must have had a beginning.  There are  more consequences of the expansion. If we run the ``movie'' of the expansion backwards, we obtain that, far enough back in the past, all the content of the Universe was concentrated in a singularity (the Big Bang). Also, during expansion, the Universe cooled, so the early Universe was hot.  The hot Big Bang   scenario is extremely successful  in  explaining some key observations: the abundance of light elements and the cosmic microwave background (CMB). The CMB is discussed in section \ref{sec:CMB}. The abundance of light elements is explained by the so-called Big Bang nucleosynthesis. I will not discuss nucleosynthesis here, although it is a fascinating subject. I will just mention that primordial nucleosynthesis took place minutes after the Big Bang and that, through the hot  starting temperature and the gradual cooling of the Universe, the nuclei of elements from deuterium, helium isotopes, lithium and beryllium were made. Elements heavier than that were subsequently made by stars.

Note that, in the  expansion of the Universe, what is expanding really is space-time. It is not that galaxies are running away from one another  through a static space-time. At first approximation, galaxies are fixed in space  and moving with it.  Galaxies in reality do move through space-time, as we will see later, but this motion is small compared to the Hubble expansion and is sourced by small inhomogeneities. As the Hubble expansion is a stretching of space-time, there is no contradiction  to find that at high redshift things seems to move at speeds comparable to the speed of light: no information is really moving through space-time at that speed (or faster).

If we consider  the expansion of the Universe and the homogeneity and isotropy properties, then it is useful to define a scale factor to describe the expansion.  The distance $r$ between any two points as a function of time $t$ can be factorized as $r(t)=r(t_0)a(t)$, where $a$ denotes the scale factor and $r(t_0)$ is the comoving coordinate. It is easy to see that the Hubble parameter is given by $\dot{a}/a$, where the dot denotes the time derivative.  Also, it is easy to see that $a=1/(1+z)$.

It is easy to conclude that, if the Universe is expanding and  its content is matter (dark or not) -- and possibly curvature -- gravity should slow down  the expansion  over time. Not only that, but by  measuring the deceleration, one should be able to ``weigh'' the Universe!  So it was a big surprise when, in 1998, two teams~\cite{refSN1,refSN2}  analysing their data  came to the conclusion  that  the expansion was actually accelerating. Their measurements  relied on surveying   a particular type of exploding  star and measuring their redshifts. These exploding stars are ``standard candles'', i.e.\ they all have the same intrinsic luminosity.\footnote{To be completely precise, these are ``standardizable'' candles; they all have approximately the same intrinsic brightness, but any variation correlates with other observables and therefore can be effectively corrected for.} The redshifts give the recession velocity, and the observed  brightness of the standard candle (the intrinsic luminosity of which is known, and the same for all standard candles) gives an estimate of the distance. Thus, one can map the recession velocity as a function of distance, and trace the expansion history of the Universe.

The principal investigators of the two teams received the Nobel Prize in Physics in 2011. The discovery of the accelerated expansion has changed the course of cosmology, with  most major observational efforts today being devoted to understanding the acceleration.

As an  interesting  historical note, recall that Einstein added a cosmological constant term ($\Lambda$) to his equations motivated by the need to make the Universe static. When Hubble   discovered the expansion of the Universe, Einstein deemed $\Lambda$ the biggest blunder of his life. However,   the discovery of the  accelerated expansion  brought $\Lambda$ back. Matter (which is pressureless if it is dark matter or has a positive pressure) makes the expansion slow down; a component with negative pressure (such as a cosmological constant) is needed to make the expansion accelerate.   So, in the standard cosmological model, there is a component to $\Omega$ made by baryonic matter ($\Omega_{\rm b}$), a component made by dark matter  ($\Omega_{\rm dm}$), a component made by radiation, which is unimportant in the energy budget  today, and a component given by $\Lambda$ ($\Omega_{\Lambda}$). \ced{However, other ``stuff'' than a simple cosmological constant (fluids with negative pressure, slowly rolling scalar fields, more exotic things such as deviations from general relativity) could fit the bill.}\aq{Please check the sense here -- it seems to be missing something!} All this goes generically under the name of ``dark energy''.  The Nobel Prize  presentation speech says: ``The discovery of accelerating expansion through studies of distant supernovae has thus changed our image of the Universe in an unexpected, dramatic way. We have realised that we live in a Universe which largely consists of components that are unknown to us. Understanding dark energy is a challenge for scientists all over the world.''

The evolution of the Universe is driven by these parameters  ($\Omega_{\rm m},\ \Omega_\Lambda,\ H_0$) through the Friedmann equations\footnote{See class slides.}\aq{Moved to footnote. Are these slides still available?} and the properties of the various components.
Note  that the different components of the Universe scale differently with the expansion, i.e.\ dilute differently as the Universe expands. For example, the cosmological constant does not dilute, whereas matter dilutes as $(1+z)^3$ .

 So, let us move on to introduce the parameters describing  clustering. We concentrate on clustering on very large scales, say from a few Mpc  and above, called large-scale structure.  In the standard cosmological model, all structures in the Universe   arise from small primordial perturbations  (see section \ref{sec:inflation}) that grew under gravity. No theory in cosmology will be able to predict that a particular galaxy will form in a specific place; a cosmological model will be able to predict the {\it statistical properties} of the fluctuations. The statistical tool most used is the {\it power spectrum}, which is the Fourier counterpart of the correlation function. The way to characterize the primordial power spectrum is as a power law with an amplitude (usually called $A$)\footnote{The  numerical value of the amplitude is dependent on the convention  used, i.e.\ the choice of the {\it pivot point}, which is the $k$ value where the amplitude is set.} and a slope ($n_{\rm s}$). Here the subscript ``s'' stands for scalar perturbations. In the standard cosmological model, scalar perturbations are all there effectively is, but beyond the (minimal) standard model there may be other type of perturbations (e.g.\ isocurvature, tensors, vorticity, etc., depending on how imaginative or non-standard one wants to be).

 Finally, there is one more parameter in the model, which is one of  those characterizing our ignorance: this is the ``optical depth'' parameter. It encloses information about the  epoch of formation of the first stars and how powerful they were at emitting radiation that  ionizes the gas in the Universe.

 So,  there are six parameters in the standard cosmological model, also called $\Lambda$CDM or LCDM \ced{(CDM standing for ``cold dark matter'')}.  These six parameters fit observations of the Universe from  $z=1100$ (corresponding to 380\,000 years after the Big Bang or, in the LCDM model, 13.7 billion years ago)  to today. The Universe is spatially flat, made of baryonic matter, cold dark matter, cosmological constant and radiation, which is unimportant today (but much more important in the past). One  more parameter describes the present-day expansion rate (the Hubble constant). In this model, inflation (see section \ref{sec:inflation}) generated the primordial perturbations that grew under gravity.  Two parameters characterize the primordial  perturbations: their power spectrum power-law  amplitude
 and slope. This is an extremely successful model and the values of the six basic  parameters  are known \ced{with per cent accuracy}.\aq{Please rewrite and quantify this, as \% accuracy could mean different things}

 It is interesting to note that the model was developed when the datasets available were  much smaller and of much lower quality  than those available today. But the model has survived the scrutiny of time  and  the stringent test imposed by   comparison to the new data.  This does not mean that there is no space for small deviations from this standard cosmological model, and in fact much of the  research in cosmology these days is focused on constraining deviations from  this model.
 This model leaves many open questions, such as the following: What is $\Lambda$? Is it really a constant?  What is dark matter?  Why are $\Omega_m$ and $\Omega_{\Lambda}$ comparable today?  What is the physics behind inflation? We will revisit these issues  later on.

 \section{The cosmic microwave background}
 \label{sec:CMB}

 The Universe cools as it expands -- the redshift results in a reduction of temperature (think of Wien's law). If the Universe is cool, large and expanding today, it must have been smaller and warmer in the past. This leads to the cosmic microwave background (CMB), the relict radiation from the Big Bang that pervades the whole Universe.  Also, while looking far away corresponds to looking back in time,  the CMB also gives a ``surface''  behind which we cannot see directly.

 Regular hydrogen gas lets light pass through more or less unimpeded. This is the case today, where the hydrogen gas is either cold and atomic, or very thin, hot and ionized.
 But in the early Universe, when it was much warmer, the gas would have been ionized, and the Universe opaque to light.
 As the Universe cooled, the electrons and protons ``recombined'' into normal hydrogen, and the Universe suddenly became transparent. This gives the last scattering surface: a snapshot  of the early Universe.

 As a historical note, the CMB was discovered ``accidentally''  in 1965 by two engineers (Penzias and Wilson)  at Bell Labs in New Jersey, USA, as a uniform glow across the sky in the radio.  At the same time a few miles away at Princeton University, researchers (Dicke,  Peebles,  Roll and Wilkinson) were looking for the ``primordial fireball'', which would be an unavoidable consequence of the hot Big Bang model.  It should be the blackbody emission of hot, dense gas (temperature $T\sim 3000$~K,  peak wavelength $\lambda_{\max}\sim1000$~nm) redshifted by a factor of 1000
to a peak wavelength of 1~mm and $T\sim 3$~K radiation.
 When they heard of the discovery at Bell Labs, Dicke was reported to say ``Well boys, we have been scooped!''   Penzias and Wilson received the Nobel Prize in 1978.
 Two papers (the discovery one and the theoretical explanation) by the two groups were published back to back in  the \textit{Astrophysical Journal} in 1965~\cite{PenziasWilson65, DPRW65}.

 We had to wait until 1992 for the results of the \textit{COBE} satellite to  confirm that the radiation was truly a blackbody and that small anisotropies (about one part in $10^5$) were present. The small anisotropies  correspond to the primordial fluctuations  that grew under gravitational instability to form all the structures we see today. The \textit{COBE} team was awarded the Nobel Prize in Physics in 2006.  The prize dedication mentions ``These measurements \dots\ marked the inception of cosmology as a precise science.'' In fact, accurate measurements of the CMB fluctuations  were crucial in establishing the standard cosmological models and constraining its parameters at the \ced{per cent level}.\aq{Please rewrite to quantify this, as \% level could mean different things.}

\begin{figure}[ht]
\begin{center}
\includegraphics[width=6cm]{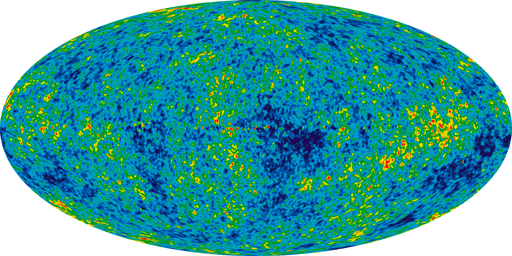}
\caption{Cosmic microwave background temperature anisotropies as seen by the \textit{WMAP} satellite. Figure courtesy of the \textit{WMAP} science team. }
\label{fig:CMBWMAP}
\end{center}
\end{figure}

 We see density perturbations as temperature fluctuations:   photons need to climb out of potential wells before they can travel to us. Thus, from  potential hills, the radiation is blueshifted and they look hotter; from  potential wells, the radiation is  redshifted and they look colder. See Fig.\ \ref{fig:CMBWMAP}   for a map of these temperature fluctuations as seen by the \textit{WMAP} satellite. It is from the statistical properties of these fluctuations  that  it was possible to  constrain cosmological parameters so well.

 Recall that the last scattering surface  (which we see as the CMB) is a snapshot of the early Universe. As the inverse was hot and photons were coupled to baryons, it gives us a snapshot of the photon--baryon fluid. On large scales, we see primordial ripples. On smaller scales, there are two competing effects: the radiation pressure from photons, and the compression from gravity. Thus sound waves are generated, which stop oscillating at recombination, and we see  in the CMB a snapshot of when the oscillations stopped.  There is another important ingredient, which is the horizon size at the last scattering surface.  The Universe back then was so dense that the speed of sound  was very close to the speed of light.  Pressure can counterbalance perturbations only on scales smaller than how far the sound could have travelled from the Big Bang to the epoch of the CMB (sound horizon). The horizon corresponds to  a fundamental mode and there should be an imprint in the CMB sound waves of    the sound horizon  as a fundamental model and its overtones.
Another interesting  fact in  the history of cosmology  (intertwined with  political history) is  the prediction of the properties of the CMB perturbations: on both sides of the Iron Curtain (Sunyaev and Zeldovich in Russia, and Peebles and Yu in the USA), the same prediction was being worked out, but the West was not aware of the developments published in Russian until they were translated (see~\cite{SZCMB,PeeblesYu}).

\begin{figure}[ht]
\begin{center}
\includegraphics[width=10cm]{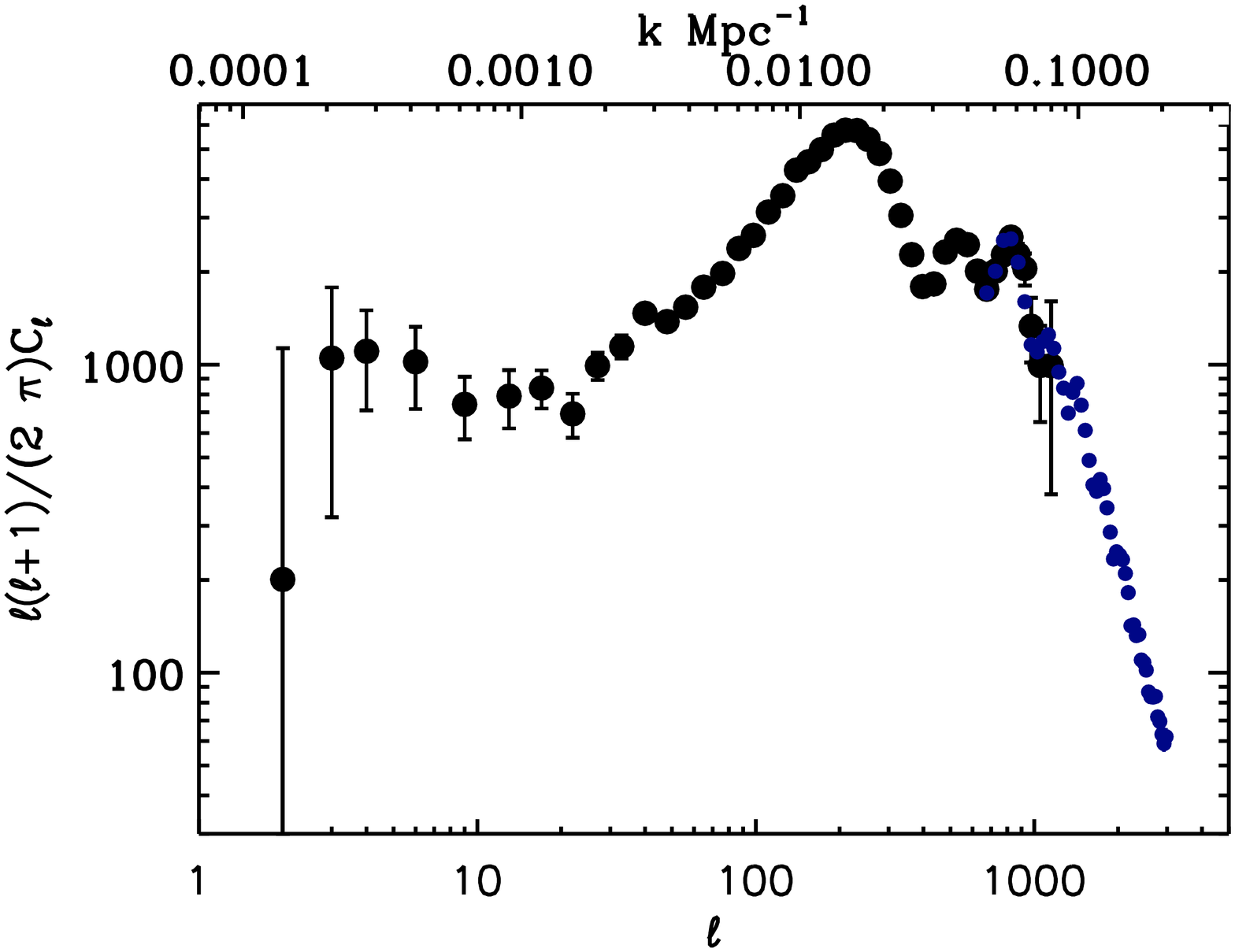}
\caption{Angular power spectrum of the cosmic microwave background temperature anisotropies as seen by the \textit{WMAP} satellite (black) and the South Pole telescope (blue points).}
\label{fig:cl}
\end{center}
\end{figure}

A CMB map such as the one shown in Fig.\ \ref{fig:CMBWMAP} must be compressed to study cosmology: the  temperature fluctuation is expanded in spherical harmonics and the temperature angular power spectrum is computed. If the anisotropy is a Gaussian random field (as  predicted by inflation, to a good approximation), then all the statistical information of the map is enclosed in the angular power spectrum $C_{\ell}$ (Fig.~\ref{fig:cl}).

Few features are evident from the $C_{\ell}$ plot. On large scales, larger than the sound horizon, we see the primordial ripples. On smaller scales, acoustic peaks are seen, which corresponds to extrema of the photon--baryon fluid oscillations. A peak is associated with the main compression mode or the fundamental tone corresponding to the sound horizon size. Smaller peaks are associated with the maximum rarefactions and compressions of the overtones. The fundamental mode acts as a standard ruler, as the sound horizon size is well determined from first principles and depends very little on cosmology.  By observing such a standard ruler, we are constructing the largest possible triangle which can then be used to determine the global geometry of space-time between us and the last scattering surface.
Also, the heights of the peaks corresponding to the compression give information about  the depth of the potential wells and thus  the total amount of matter. The relative heights of the peaks corresponding to compression (odd peaks) and rarefaction (even peaks) yield information about the amount of baryonic matter.  Finally, all these processes originated from a primordial spectrum of fluctuations and the $C_{\ell}$ also carries information on that.  (For a more rigorous derivation of all this, see e.g.~\cite{waynecmb1,waynecmb2}.)

\begin{figure}[ht]
\begin{center}
\includegraphics[width=8cm]{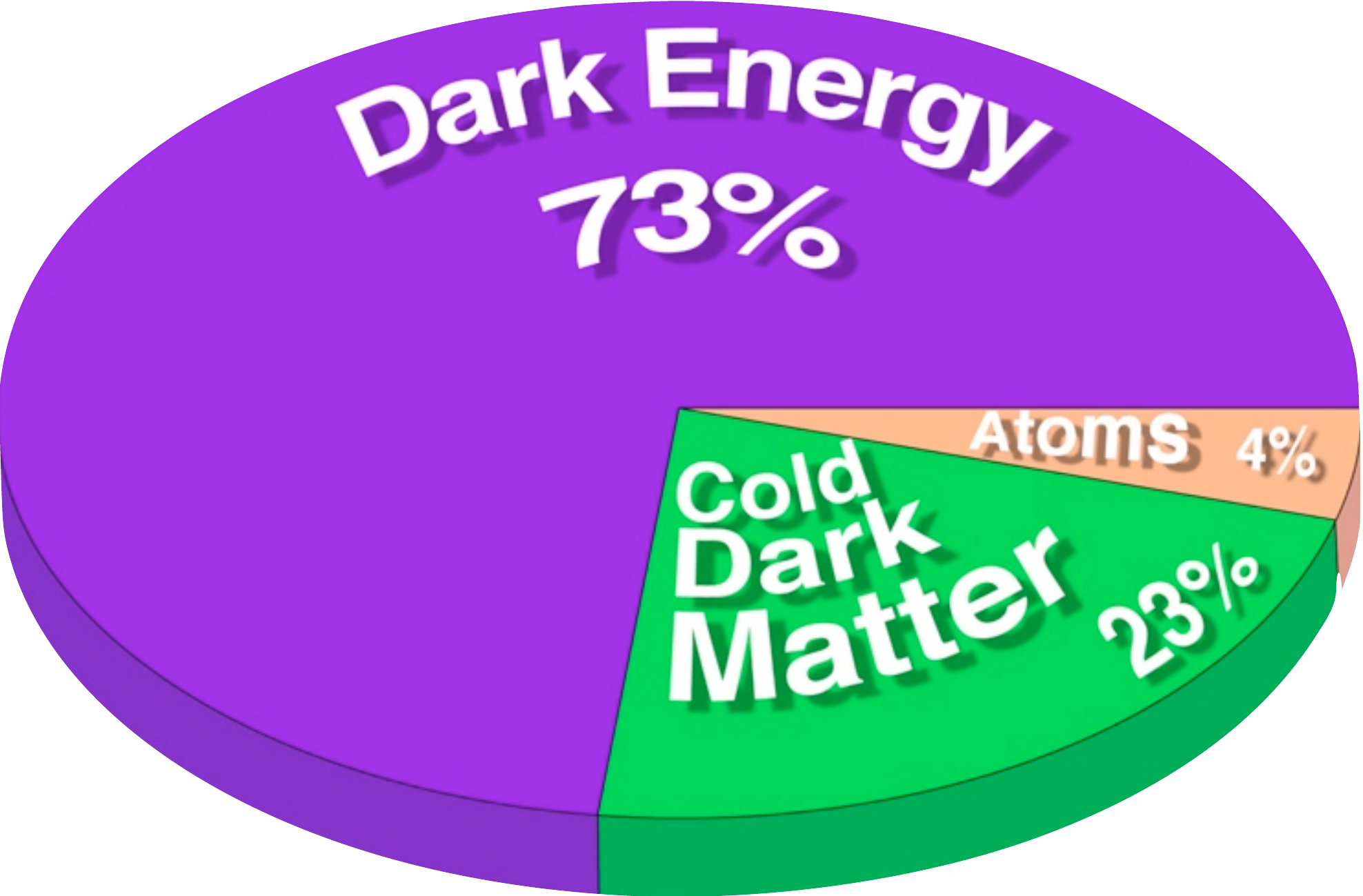}
\caption{The Universe composition in the standard model for cosmology. The CMB data from \textit{WMAP} were crucial to establish this model.}
\label{fig:cosmicpie}
\end{center}
\end{figure}

A key role in establishing the current standard model of cosmology was played by the \textit{WMAP} satellite. \textit{WMAP} produced the highest-resolution  full-sky map of the CMB anisotropies to date.
A detailed analysis of the shape of the angular temperature power spectrum  from the \textit{WMAP} data constrains the parameters of the LCDM  model at the \ced{per cent level}\aq{Please rewrite to quantify this, as \% level could mean different things.} (see~\cite{WMAP1,WMAP3, WMAP7} and references therein). For example, the composition of the Universe is summarized in Fig.~\ref{fig:cosmicpie}.

Before concluding this section, we should mention that the \textit{Planck} satellite has been measuring the CMB sky with even better precision than \textit{WMAP} and should release its results in a year or so.
There is more information in the CMB radiation than the temperature fluctuations: the CMB light is also polarized. This will be discussed in section~\ref{sec:polarization}.

 \section{Large-scale structure}
 \label{sec:lss}

 Large-scale structure denotes the large-scale distribution of matter  that is traced by galaxies (and gas). When we see the  primordial fluctuations in the CMB (albeit with the processing due to the fact that we see the radiation, i.e.\ processed through the {\it radiation transfer function}), we see the initial conditions. When we observe the large-scale structure, we see the result of the action of gravity. Of course, if we trace large-scale structure using the galaxy distribution (or the intergalactic hydrogen gas via observations of the so-called Lyman-alpha forest), we must bear in mind that galaxies may not be perfectly faithful tracers of the dark matter distribution, i.e.\ they could be {\it biased} tracers. On the other hand, it has recently become possible to trace the dark matter distribution directly via gravitational lensing. At present, gravitational lensing measurements on large-scale structure scales still  have large  associated error bars, but in the near future specifically designed surveys will make gravitational lensing  an invaluable tool to map the large-scale structures in an unbiased way.

Again, cosmology deals with statistical  properties of the large-scale structure, and the most widely used tools are the power spectrum and the corresponding two-point correlation function. Note that, even if the initial conditions were perfectly Gaussian, nonlinear gravitational evolution would make the distribution non-Gaussian,
\footnote{In fact, if $\delta=\delta \rho/\rho$, where $\rho$ stands for the density, in a Gaussian distribution  the average value of $\delta$ is zero and the distribution is symmetric around zero. But as perturbations grow under gravity, overdense regions can grow almost without limit (e.g.\ you, as a human being with an average density close to the one of water, do correspond to $\delta \sim 2 \times 10^{30}$), but underdense regions can never go below $\delta=-1$; therefore, a skewness develops.}\aq{Is the sense in the footnote OK as slightly rewritten?}
 and thus the power spectrum of large-scale structure does not include all the statistical properties of the distribution. Nevertheless, the power spectrum of large-scale structure is a powerful tool to constrain cosmology. Similarly to the CMB transfer function, the primordial spectrum also gets processed through the so-called matter transfer function. I will not go into further details here, but I will just say that, for a given model, the matter transfer function can be computed very accurately and thus there is (within a model) a one-to-one correspondence between the primordial spectrum and the matter power spectrum. To date, the power spectrum has been probed from scales of $k=0.001~h$/Mpc to highly nonlinear scales of $\sim 5~h$/Mpc using different tracers and different measurement approaches.  The standard cosmological model as constrained by, for example, CMB-only observations fits extremely well the large-scale structure observations.
The combination of the two datasets can be used to break some parameter degeneracies but, more importantly, to test for deviations from the standard model.

The CMB data alone  in particular are not too stringent in measuring the value of  $\Omega_{\Lambda}$ once the spatial  flatness  assumption is dropped   or even imposing flat geometry if constraining the properties of dark energy (whether it is  truly a cosmological    constant or some dynamical quantity). The limitation of the CMB in constraining dark energy  is that the CMB is situated at a given redshift ($z=1090$ in the standard cosmological model  to be precise). As discussed in section \ref{sec:DE}, dark energy is seen through the acceleration of the expansion rate, so one really needs to be able to look at more than a single snapshot of the Universe.  The large-scale structure allows one to do that. In fact, the  fundamental mode that we see in the CMB, given by the sound horizon at last scattering, is also imprinted in the large-scale structure clustering. In fact, imagine a perturbation with dark matter, baryonic matter and photons starting all in the same phase. Photons freely stream out of the dark matter potential well carrying the baryons with them. At decoupling, baryons find themselves offset with respect to the dark matter and fall into the dark matter potential wells. But, as baryons are one-seventh of the dark matter, the dark matter ``feels'' the baryons also through gravity. Thus the   same acoustic oscillations present in the CMB radiation should be present in the dark matter distribution, albeit with a much smaller amplitude. If, then, galaxies form  in the dark matter potential wells, the galaxy distribution should also carry the same imprint. This signature goes under the name baryon acoustic oscillations (BAO)~\cite{EisensteinBAO} and was first detected  in galaxy surveys in 2005~\cite{baodetection1, baodetection2}. Of course, if the BAO signature can be detected in a large-scale structure survey  at different redshifts, one would have a standard ruler  with which to trace the expansion history of the Universe.
More precisely, a measurement of the radial BAO signal would give a combination of the standard ruler and the Hubble parameter as a function of redshift, and the angular BAO signal would yield a combination of the standard ruler and the angular diameter distance. This is what current, forthcoming and planned surveys are aiming to measure.
In addition, the shape of the large-scale structure power spectrum encloses information not only about the primordial fluctuations but also about  cosmological parameters.
Of course, there are many challenges in this measurement beside the technical difficulty of surveying  the position of galaxies over a sizeable part of the observable Universe; among them, modelling the effects of nonlinear evolution, and the fact that galaxies may not be faithful tracers of the underlying dark matter distribution.
To date, the most recent  constraints on cosmology from such a signal are presented in Refs.~\cite{BethDR7,WillDR7,wiggle-z}.

\begin{figure}[ht]
\begin{center}
\includegraphics[width=7cm]{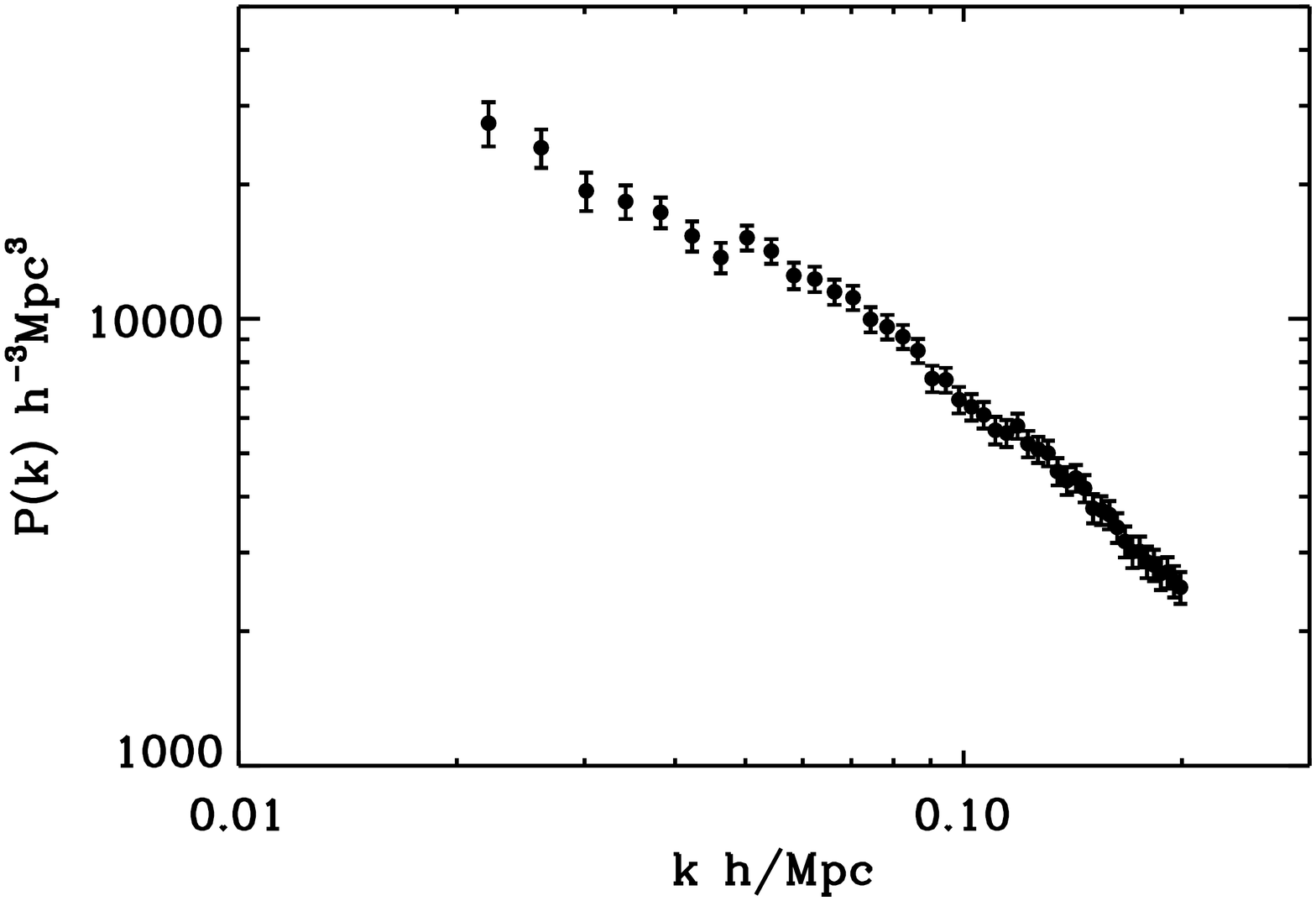}
\includegraphics[width=7.2cm]{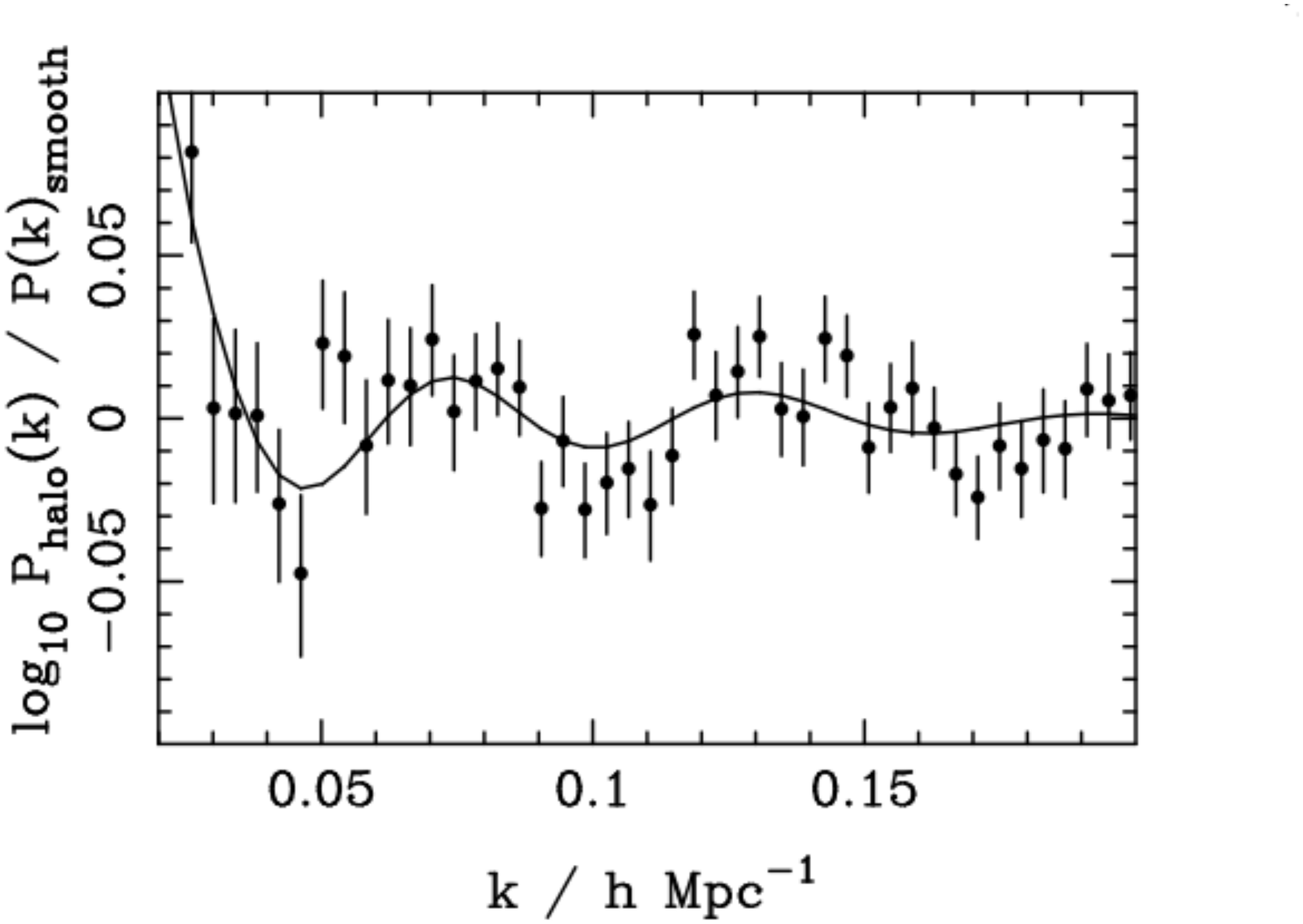}
\caption{(left)~The large-scale structure power spectrum as measured from the Digital Sky Survey data release~7. (right)~The same power spectrum divided by a smooth one to highlight the BAO feature. The solid line is the theory prediction for a standard LCDM model. Figure reproduced from Ref.~\cite{BethDR7}.}
\label{fig:LSS}
\end{center}
\end{figure}

One  aspect  that the combination of CMB and large-scale structure can measure well and   for which  future data promise  to be particularly constraining  and of interest to this audience is neutrino properties.

\section{Neutrino properties and cosmology}

 Neutrino oscillations indicate that neutrinos have mass, although neutrino mass eigenstates are not the same as flavour eigenstates. The standard model has three neutrinos and, combined with oscillation constraints  (one  mass  splitting much larger than the other one), one concludes that three  neutrino mass hierarchies are possible: normal (one heavy neutrino and two light ones -- NI), inverted (two heavy and one light -- IH) and degenerate (where the mass splitting is small compared to the overall neutrino mass scale).
 While particle physics experiments are extremely sensitive to flavour but much less sensitive to the absolute mass, cosmology is insensitive to flavour but sensitive to the absolute mass scale.
 Cosmological constraints on neutrino properties are highly complementary to particle physics experiments for several reasons, as we explain below.
 
 \begin{enumerate}

\item\textbf{Relict neutrinos.}\\ 
Relict neutrinos produced in the early Universe are hardly detectable by weak interactions, making it impossible with foreseeable technology to detect them directly. But  cosmological probes  offer the opportunity to detect (albeit indirectly)  relict neutrinos. The hot Big Bang model predicts  a background of relict neutrinos in the Universe with an average number density of $\sim 100 N_{\nu}$~cm$^{-3}$. These neutrinos decouple from the CMB at redshift $z\sim 10^{10}$ when the temperature was about a few~MeV, but remain relativistic down to much lower redshifts depending on their mass.

\quad Massive neutrinos affect cosmological observations in different ways.
Primary CMB data  alone  can constrain the total neutrino mass, $\Sigma$,  if it is above $\sim 1$~eV (\cite{WMAP7} find $\Sigma<1.3$~eV at 95\% confidence) because these neutrinos become non-relativistic before recombination, leaving an imprint in the CMB. Neutrinos with masses $\Sigma<1$~eV become non-relativistic after recombination, altering matter--radiation equality for fixed $\Omega_m h^2$; this effect is degenerate with other cosmological parameters from primary CMB data alone.
After neutrinos become non-relativistic, their free streaming damps the small-scale power and modifies the shape of the matter power spectrum below the free-streaming length. The free-streaming length of each neutrino family depends on its mass.

\quad Current cosmological observations do not detect any small-scale power suppression and break many of the degeneracies of the primary CMB, yielding constraints of $\Sigma<0.3$~eV~\cite{BethDR7}. A detection of such an effect, however, would provide a detection, although indirect, of the cosmic neutrino background. The fact that  oscillations predict a  minimum total mass $\Sigma \sim 0.54$~eV implies that  forthcoming surveys  have the  statistical power to detect the cosmic neutrino background.

\item\textbf{Cosmology remains a key avenue to determine the absolute neutrino mass scale.}\\
Particle physics experiments will be able to place lower limits on the {\it effective} neutrino mass which depend on the hierarchy, with no rigorous limit achievable in the case of normal hierarchy. Neutrinos free streaming suppress the small-scale clustering of large-scale cosmological structures  by an amount that depends on neutrino mass. Planned surveys  should yield an error on $\Sigma$ of  0.04 (for details  see Ref.~\cite{Carbone/etal:2011}) and therefore will not only  detect the effect of massive neutrinos on clustering but also  determine the absolute neutrino mass scale.

\item\textbf{``What is the hierarchy (normal, inverted or degenerate)?''}\\
Neutrino oscillation data are unable to resolve whether the mass spectrum
consists of two light states with mass $m$ and a heavy one with mass $M$ (NH) or two heavy states with mass $M$ and a light one with mass $m$ (IH) in a model-independent way. Cosmological observations  can help to determine the hierarchy.
Since cosmology is insensitive to flavour, one might expect that cosmology may not help in determining the neutrino mass hierarchy. However, for $\Sigma <0.1$~eV, only normal hierarchy is allowed, and thus a mass determination  can help disentangle the hierarchy. There is, however, another effect: neutrinos of different masses become non-relativistic at slightly different epochs; the free-streaming length is slightly different for the different species, and thus the detailed shape of the small-scale power suppression depends on the individual neutrino masses (not just on their sum).

\quad As discussed in Ref.~\cite{JKGV10}, in cosmology one can safely neglect
the impact of the solar mass splitting. Thus two masses characterize the neutrino mass spectrum, the lighter one, $m$, and the heavier one, $M$.  The mass splitting can be parametrized by  $\Delta= (M-m)/\Sigma$ for normal hierarchy and $\Delta=(m-M)/\Sigma$ for inverted hierarchy. Cosmological data are  very sensitive to $|\Delta|$; the direction of the splitting (the sign of $\Delta$) introduces a subdominant correction to the main effect. Planned surveys should be able to measure $|\Delta|$ and may be able to help to determine the hierarchy (i.e.\ the mass splitting and its sign)  if far enough away from the degenerate hierarchy (i.e.\ if $\Sigma<0.13$).

\item\textbf{``Are neutrinos their own antiparticle?''}\\
If the answer is ``yes'', then neutrinos are Majorana fermions; if not,
they are Dirac. If neutrinos and antineutrinos are identical, there could have been a process in the early Universe that affected the balance between particles
and antiparticles, leading to the matter--antimatter asymmetry we
need to exist.  This question can, in principle, be resolved if  neutrinoless double-beta decay is observed. But if  such experiments  lead to a negative result, the implications for the nature of neutrinos depend on  the hierarchy.
As shown in Ref.~\cite{JKGV10} (and references therein), in this  case cosmology can offer complementary information by  helping to determine the hierarchy.

\item\textbf{Number of neutrino species.}\\
After decoupling, neutrinos contribute to the relativistic energy density.
Cosmology is sensitive to the physical energy density in relativistic particles in the early Universe $\omega_{\rm rel}$, as it affects the expansion history; $\omega_{\rm rel}$ includes photons and neutrinos $\omega_{\rm rel}=\omega_{\gamma}+N_{\rm eff}\omega_{\nu}$, where $\omega_{\gamma}$ is the  energy density in photons, $\omega_{\nu}$ the energy density in one  neutrino and $N_{\rm eff}$ is the effective number of neutrino species. In the standard model with three neutrinos $N_{\rm eff}=3.046$ to account for quantum chromodynamic effects  and for neutrinos being incompletely decoupled during electron--positron annihilation.

\quad With $\omega_{\gamma}$ being extremely well constrained from the measurement of the CMB  temperature, constraints on $\omega_{\rm rel}$ can be interpreted in terms of $N_{\rm eff}$, although a deviation from $N_{\rm eff}=3.046$ may indicate not only extra neutrino species but also  any process that affects the expansion history. Any  light particle that does not couple to electrons, ions and photons will act as an additional relativistic species. \ced{The relativistic energy density}\aq{Added this text to avoid starting sentence with a lower case symbol. OK?} $\omega_{\rm rel}$ (and therefore $N_{\rm eff}$) impacts the Big Bang nucleosynthesis through it effect on the expansion rate. The relativistic energy density also affects the CMB via the expansion rate, which alters the  epoch of matter--radiation equality.

\end{enumerate}


 \section{Inflation }
 \label{sec:inflation}

 The \ced{popular}\aq{Better than ``hit''} Big Bang model (combined with general relativity and the cosmological principle) is extremely successful at predicting observations such as Hubble's law, the CMB and the abundance of light elements. However, it leaves some major open problems:  the flatness problem, the horizon problem and the monopole problem  (and the origin of the cosmological perturbations). In other words, \ced{we still have the following puzzles:}\aq{I think this needed more of an introduction, so added this because ``puzzles'' is also mentioned by you below. OK?} 
 
\begin{enumerate}
\item[(a)] 
Why is the Universe so flat? The Friedmann equations extrapolated back in time\footnote{See class slides.}\aq{Moved to footnote. Are these slides still available?} basically  say  that,  in order for $|\Omega-1|< 0.1$, say, as observations indicate, very early on $|\Omega-1| < 10^{-60}$.
\item[(b)]
Why is the Universe so big? The Universe is homogeneous on large scales. Take, for example, the CMB radiation: two antipodal points are separated by 30\,000~Mpc and are both at 2.726~K (modulo the CMB temperature fluctuations, which are, however, very very small), but the age of the Universe back then was only 380\,000 years. Light could have travelled only  380\,000 light years, i.e.\ only points closer than that (i.e.\ closer than about 0.2~Mpc) could have been in \ced{causal contact.}\aq{Here and below, changed ``casual contact'' to ``causal contact'' OK?} How, then, can two antipodal points be at the same temperature? 
\item[(c)]
If the Universe cooled from an extremely hot phase, in the very early \ced{Universe} there  should have been phase transitions and they should have left cosmological defects. Of all the types of cosmological defects, monopoles are the most dangerous ones: more than one in the observable Universe would  be catastrophic, i.e.\ would   violate all observations.
\item[(d)]
Finally, where do the primordial perturbations we talked about before come from?
\end{enumerate}

 The paradigm of {\it inflation} comes to the rescue. \ced{The concept of inflation} was started in 1981 by Alan Guth, but  it is still a very active area of research and is still a paradigm rather than a specific model or theory. Inflation postulates a brief period of accelerated expansion in the very early Universe, something not too dissimilar from  a cosmological constant.  In particular, one postulates the existence of one (or more) field (the inflaton) which rolls down  a  potential  very slowly. Specific models have different shapes for the \ced{inflaton potential.}\aq{Please check terminology here and elsewhere -- is it ``inflation potential'' or ``inflaton potential''?}

 This solves all the above puzzles in one go.  (c)~Monopoles are diluted away very quickly so that there would be no more than one in the observable Universe.
 (a)~The flatness problem is solved, as a rapid expansion by many e-foldings would make any non-flat geometry arbitrarily close to flat. 
 (b)~In order to understand the solution of the  horizon problem, we need to realize that in cosmology there are different types of horizons and this happens because light travels at finite speed (which gives a horizon) but at the same time the Universe is expanding. Thus the {\it particle horizon} is the maximum comoving distance light could have propagated in a  given interval of time   from a specific $t_1$ to $t_2$, and takes into account the past expansion history. Note that, for the same time interval, the particle horizon can be very different if the expansion history changes!  Then there is the Hubble horizon, which is just the speed of light divided by the time interval. For normal expansion histories, the Hubble horizon is not too different from the particle horizon, but for weird expansion histories the two can be very different. For example, in an   accelerated expansion, the particle horizon becomes much bigger than the Hubble horizon. So all regions of the Universe, even if  separated by many Hubble horizons, are so uniform because in the past they were in \ced{causal contact} -- they are all within each other's particle horizon!  \ced{Because of (b), inflation generated super-horizon perturbations (of course, this refers to the Hubble horizon).}\aq{Moved to here rather than being separate one-sentence paragraph after (d). OK?}
(d) As an added bonus inflation generates Gaussian perturbations:
quantum fluctuations stretched to become classical
by the expansion. The mechanism is similar to Hawking radiation. (The event horizon being the complementary concept to the particle horizon.) It turns out that the power spectrum of these perturbations will be a power law: different models of inflation predict slightly different power laws, but all are close to ``scale-invariant''.

 One may ask, can this paradigm \ced{of inflation} be tested?
 Inflation predicts a flat Universe, and so far observation says that the Universe is consistent with being flat. Inflation predicts  a power-law primordial power spectrum close to scale-invariant, and specific models yield  slightly different slopes. Observations are fully consistent with a  power-law power spectrum  close to scale-invariant, and we hope to be able to distinguish different inflationary models by determining the slope accurately. For example,  current data give $n_s=0.968 \pm 0.012$~\cite{WMAP7}, which  excludes the perfectly scale-invariant spectrum at the 99.5\% confidence level.
 Inflation predicts   primordial perturbations very close to Gaussian, and indeed this is confirmed by observations.
 Inflation predicts super-horizon perturbations and a stochastic background of gravity waves. These prediction can also be tested with observations. To understand this, we first need to introduce CMB polarization.

 \subsection{CMB polarization}
 \label{sec:polarization}
 
 The CMB light was predicted to be polarized shortly after the CMB was discovered. The CMB polarization signal was first detected  on a patch of the sky by the DASI experiment in 2002~\cite{dasi1,dasi2} and then mapped over the full sky by \textit{WMAP}.

Temperature quadrupole at the last scattering surface generates polarization via \ced{Thomson scattering}.\footnote{See class slides.}\aq{Moved to footnote. Are these slides still available?} On sub-horizon scales it is the density perturbations that give the quadrupolar pattern. One should therefore expect a radial pattern around a hot spot and a tangential one around cold spots. This has been seen (albeit by stacking the pattern around many spots) in 2010 (see Ref.~\cite{WMAP7}). For density  fluctuations on large (super-horizon scales) it is the velocity they source (from hot spots to cold spots)  that generates the quadrupolar pattern.
Since the velocity is out of phase with the density perturbation, and on super-horizon scales the temperature signal is sourced by the density and the polarization by the velocity, one should expect an anticorrelation between temperature and polarization on these scales~\cite{ZaldarriagaHarari, SpergelZaldarriaga97}, which has been seen (see Ref.~\cite{Peiris/etal:2003}). This is the signature of  the existence of super-horizon perturbations as predicted by inflation.

There is more. Density perturbations have no \ced{handedness}\aq{Is this terminology OK -- not ``handiness''} and therefore the polarization perturbations also do not have \ced{handedness} (recall: radial or tangential patterns around spots). Gravity waves stretch space-time and also generate a quadrupolar pattern at the last scattering surface. But gravity waves are tensor perturbations and  have a \ced{handedness}.
The CMB polarization pattern can be decomposed into two modes in analogy to electromagnetism: the E mode (which has no \ced{handedness}) and the B mode (with \ced{handedness}). The B-mode pattern cannot be generated by scalar perturbations but can be generated by  gravity waves. A detection of primordial B-mode polarization in the CMB would be a detection of the inflationary stochastic gravity wave background.  This is the holy grail of CMB polarization.

The measurement of CMB primary polarization is, however, very difficult. In fact, the signal is very small (orders of magnitude below the temperature signal) and is contaminated by the emission of our own Galaxy.

In  summary, quantum fluctuations get stretched to become classical and ``super-horizon'' because of the accelerated expansion.
The spacing of the fluctuations (their power as a function of scale)
depends on how fast they exited the horizon (i.e.\ on the Hubble parameter during inflation), which in turn depends on the \ced{inflaton potential}. Thus the shape of the primordial power spectrum
encloses information on the shape of the \ced{inflaton potential}!
In general, the observational constraints of \ced{$N_{\rm efold}>50$}\aq{OK as italic with subscript?} requires the potential to be flat (not every scalar field can be the inflaton). Detailed measurements of the shape of the power spectrum can rule in or out different potentials.
So far, observations are  consistent with the simplest inflationary models; from  current datasets, some  specific inflationary models are being critically tested.

For more information and an overview of references and possible forthcoming data, see Refs.~\cite{CMBtaskforce, BaumannWP} and references therein.

\section{Dark energy}
\label{sec:DE}

If we go back to look at Fig.~\ref{fig:cosmicpie} we quickly realize that in the standard cosmological model  96\% of the Universe is missing, i.e.\ it is supposed to be  in the form of dark matter or dark energy, about which we know very little. While the dark matter at least is a form of matter and may soon be detected directly by one of the underground ongoing experiments, there are two major open questions  of the model that can be  observationally addressed exclusively by looking at the sky: What created the primordial perturbations? What makes the Universe accelerate?  These two questions may not be unrelated.    The first question was examined in section~\ref{sec:inflation}, so here we concentrate on the second question.

A major part of the observational effort in cosmology today is devoted to the accelerating Universe and  dark energy.
It is important to stress a peculiarity of cosmological observations: this sets a fundamental limit on how well a property of the Universe or a parameter of the model  can be measured, which is called the cosmic variance limit.
The cosmological principle of homogeneity and uniformity  implies that we can use statistical properties of the Universe as observables and as a measurable quantity to make connection with theoretical predictions. Inflation also tells us that the visible Universe is only a small part of the Universe. It is only possible to observe part of the Universe at one particular time,
so it is difficult to make statistical statements about cosmology on the  scale of the entire Universe, as the number of  independent observations (sample size) is finite.  In cosmology we are interested in the properties of the underlying model, which is valid for the entire Universe, not just the observable Universe. This sets an error floor, and a fundamental limit to any measurement, the so-called cosmic variance error.

Like every cloud has a silver lining, this means that  in principle in cosmology one can plan the perfect experiment: repeating the experiment or improving it will not improve the statistical errors. In cosmology, therefore, we can do what I call ``ultimate experiments''.  The interesting new development  is that  planned and some ongoing surveys are  ultimate experiments.

Back to  dark energy. If we assume a model where the Universe is composed by matter and a cosmological constant, and we impose spatial flatness, then CMB observations alone indicate that $\Omega_m \sim 0.27$ and $\Omega_{\Lambda} \sim 0.73$. If we do not impose flatness, then we need to consider along with the CMB also  supernovae data or large-scale structure data or  measurements of the Hubble parameter (or more than one of these,  of course). In this case, then,  a flat model  with the above values for matter and cosmological constant is a very good fit to the observations.

In Einstein's equations, the cosmological constant can be interpreted as vacuum energy.
However, when computing the expected vacuum energy density from the quantum field theory approach, the theoretically predicted value exceeds the measured one by \ced{123 orders of magnitude.}\aq{Is this the correct value?} This has been dubbed the ``cosmological constant problem''.
The vacuum energy density  (which does not dilute as the Universe expands) would have made a negligible  contribution in the early Universe but  the future Universe would grow so quickly as to be essentially empty of matter. This is the ``Why now?'' problem: Why  is the brief epoch when the cosmological constant is non-negligible but still comparable with the matter density  coincident with the present day?  Why is the present day so special?

\begin{figure}[ht]
\begin{center}
\includegraphics[width=10cm]{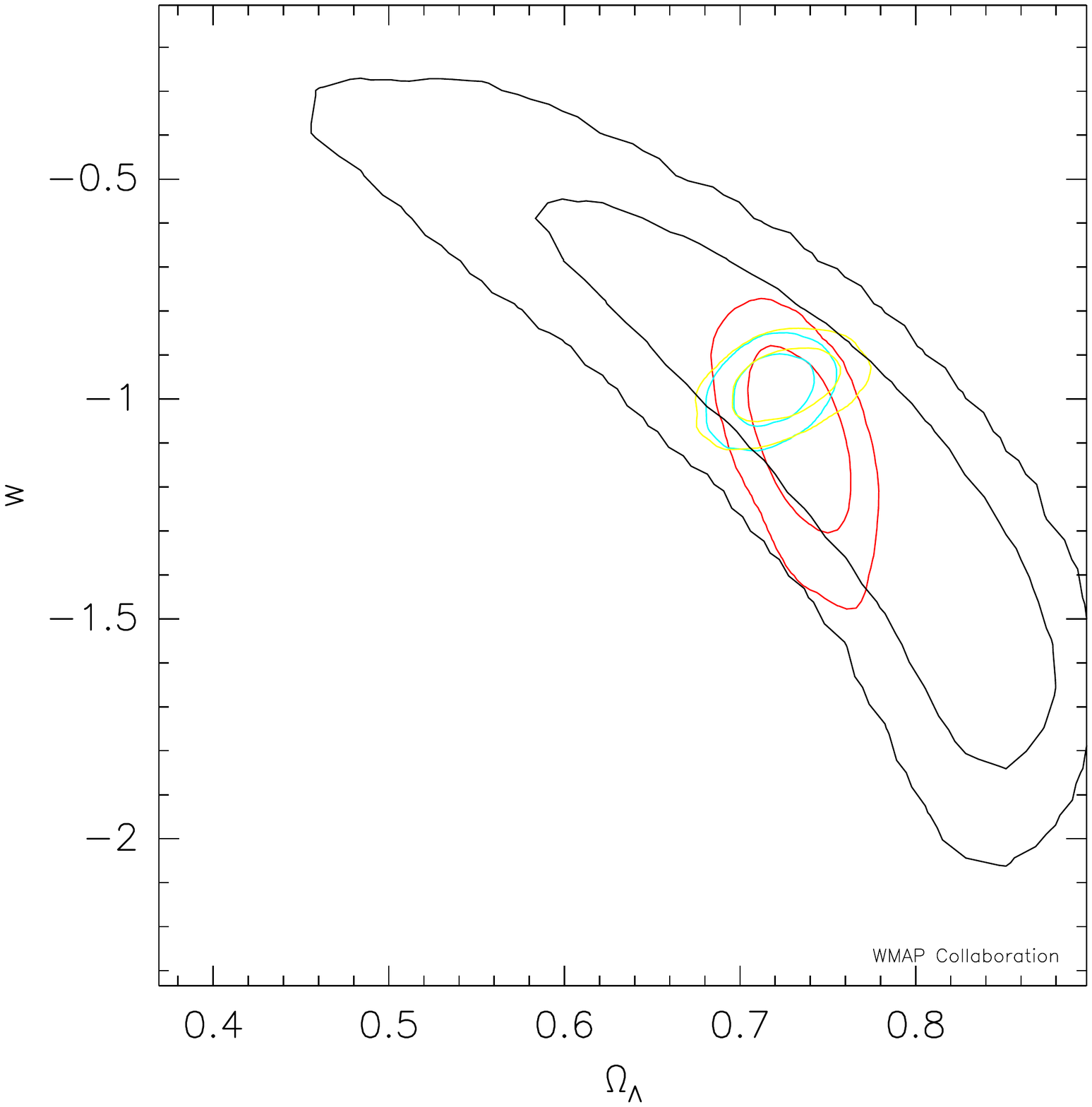}
\caption{Observational constraints on the  dark energy equation-of-state parameter assumed to be constant with redshift and assuming a flat Universe geometry: black, \textit{WMAP}, seven years only; red, \textit{WMAP} + BAO + $H_0$; yellow, \textit{WMAP} + supernovae; cyan, \textit{WMAP} + BAO + supernovae.}
\label{fig:w}
\end{center}
\end{figure}

To overcome these problems, one  needs to look for different physical mechanisms. A dynamical option is to suppose that a cosmic scalar field, minimally coupled to gravity, called ``quintessence'' , changing with time and  slowly varying across space, is slowly approaching its ground state. In the basic quintessence scenario, the dark energy enters only at late times in the evolution of the Universe. More realistic scalar field models of quintessence track the dominant component of the cosmological cosmic fluid until the current epoch, when the quintessence energy overtakes the matter density.
In the  case of quintessence, dark energy can be interpreted as a fluid with an effective equation-of-state parameter $w=p/\rho$, where $p$ demotes the pressure. For a cosmological constant $w\equiv -1$, for quintessence $w\ne -1$ and it can be a function of redshift. Current constraints on $w$ assumed to be constant in redshift and assuming a flat Universe are shown in Fig.~\ref{fig:w}. Note that, so far, observations are  consistent with a cosmological constant.

It is customary to parametrize a possible evolution of the equation-of-state parameter as a linear function of the scale factor. This parametrization is useful to quantify the performance of future surveys or to quantify how different from a cosmological constant dark energy may be as long as observations give back a $w$  consistent with $-1$, but it does not have any physical motivation.

The conclusion   that we live in such a strange Universe relies on the assumption that general relativity holds on scales comparable to the horizon.  In fact the evidence for dark energy comes from observations of the Universe on extremely large scales, comparable to the size of the observable Universe. However, precision tests of gravity have been made only up to Solar System scales. An enormous extrapolation (13 orders of magnitude) is needed to apply it to horizon scales.  It may be that gravity as we know it breaks down on those scales, and  we interpret that as an accelerated expansion.
 Is it possible to distinguish between all these different possible explanations for dark energy (i.e.\ cosmological constant, a slowly varying scalar field or modifications of general relativity on large scales)?

 Let us review the observational effects of dark energy. They can be divided into two broad classes. The first is observations of the expansion history of the Universe. These are  the recession velocity versus brightness of standard candles (e.g.\ supernovae), the position of the CMB acoustic peaks, which gives the angular diameter distance to the last scattering surface, and the BAO, which give  angular diameter distance and Hubble parameter as a function of redshift. The second is the growth  rate of structures.
 The growth of cosmological perturbations is driven by gravity but its rate is a result of two competing effects: the pull of gravity and the  expansion rate of space-time. If gravity is well described by  GR, then  the growth rate of structure  can yield unequivocally the expansion rate and thus the properties of a cosmological constant or quintessence.
 Galaxy (or gravitational lensing) surveys can yield a measurement of the  growth of structure because they  can measure  the   amplitude of the power spectrum as a function of redshift.
 The clustering of galaxies, however, may not be a faithful tracer of the density field, as galaxies may be biased tracers. In fact, the process of galaxy formation is complex and may not be solely driven in a deterministic and simple way  by the dark matter overdensity.

On the other hand, galaxy surveys also probe  the peculiar  velocity field. In fact, galaxy surveys  use the observed galaxy  redshift as a distance indicator; this would be a faithful tracer of the distance in a perfectly uniform Universe where everything moved with the Hubble flow. In a clumpy Universe, the Hubble flow is disturbed by overdensities and so clustering in the line-of-sight  direction gets distorted. While this may seem a nuisance, if it can be correctly modelled and interpreted, it is not. In fact, such distortions directly probe the velocity field. The velocity field is sourced directly by the potential, thus it is a direct probe of the density perturbations and it is not as subject to bias as is the galaxy overdensity field.

On the other hand, if gravity on large scales deviates from general relativity, this would affect both the expansion history and  the growth rate of structure. Any expansion history (due to quintessence or due to modifications of gravity) can be interpreted as a quintessence model with a suitable $w(z)$. But if gravity is modified, then there would be a mismatch between expansion history and growth.
This is the reason why forthcoming surveys are poised to measure both expansion history and growth of structures -- see e.g. Refs.~\cite{DETF,Peacockreport} and references therein.

\section{Conclusions}

The standard cosmological model is extremely successful: with a handful of parameters it can describe observations of the Universe on large scales from 380\,000 years after the Big Bang until today (13.7 billion years after the Big Bang). The parameters of the model are constrained observationally with \ced{per cent precision.}\aq{Please rewrite and quantify this, as \% precision could mean different things} However, we have no real understanding of what  these parameters mean.
 Cosmology is now in a similar stage in its intellectual development to particle physics four decades ago when particle physicists converged on the current Standard Model. The Standard Model of particle physics fits a wide range of data, but does not answer many fundamental questions, for example: What is the origin of mass? Why is there more than one family?  Similarly, the standard cosmological model has many deep open questions: What is the dark energy? What is the dark matter? What is the physical model behind inflation (or something like inflation)?
 The questions address an emerging model of the Universe that connects physics at the most microscopic scales to the properties of the Universe and its contents on the largest physical scales.

It is interesting to note that, while dark matter may soon be detected directly by underground experiments, the open questions of dark energy and inflation can  be addressed almost exclusively by observing the cosmos.  Dark energy and inflation both rely on an accelerated expansion of the Universe.

Despite inflation happening 13.7 billion years ago and dark energy happening
today,  we seem to know much less about dark energy. In fact, we can test  about 10 efoldings of inflation  by looking at cosmological structures, but we cannot see dark energy  perturbations and we can only see about two efoldings.
But we can follow the (recent) expansion history and the growth of
cosmological structures  and this is what forthcoming surveys are poised to do. The wealth of forthcoming cosmological data will provide ever more rigorous tests of the cosmological standard model and search for new physics beyond the Standard Model.

\section*{Acknowledgements}

We wish to thank  the organizers of the  2011 CERN--Latin-American School of High-Energy Physics in Natal, Brazil, and all the participants and lecturers for a memorable week.  LV acknowledges support  of  FP7 IDEAS-Phys.LSS 240117.




\begin{thebibliography}{99}

\bibitem{refbullet} 
D. Clowie \textit{et al.}, \textit{Astrophys. J.} {\bf 648} (2006) L109--L113.

\bibitem{NFW} 
J.F. Navarro, C.S. Frenk and S.D.M. White, \ced{\textit{Astrophys. J.}}\aq{All of the journal abbreviations have been changed to short title form. Please check that they are all correct} {\bf 490} (1997) 493.

\bibitem{refSN1} 
A.G. Riess, A.V. Filippenko, P. Challis, \textit{et al.}, \textit{Astron. J.} {\bf 116} (1998) 1009.

\bibitem{refSN2} 
S. Perlmutter \textit{et al.}, \textit{Astrophys. J.} {\bf 517} (1999) 565.

\bibitem{PenziasWilson65} 
A.A. Penzias and R.W. Wilson, \textit{Astrophys. J.} {\bf 142} (1965) 419.

\bibitem{DPRW65} 
R.H. Dicke, P.J.E. Peebles,  P.G. Roll and D.T. Wilkinson,  \textit{Astrophys. J.} {\bf 142} (1965) 414.

\bibitem{SZCMB} 
R.A. Sunyaev and Y.B. Zeldovich,   \textit{Astrophys. Space Sci.} {\bf 7} (1970) 3.

\bibitem{PeeblesYu} 
P.J.E. Peebles and J.T. Yu \textit{Astrophys. J.} {\bf 162} (1970) 815.

\bibitem{waynecmb1} 
W. Hu and N. Sugiyama, \textit{Astrophys. J.} {\bf 444} (1995) 489.

\bibitem{waynecmb2} 
W. Hu and N. Sugiyama, \textit{Phys. Rev.} {\bf D51} (1995) 2599.

\bibitem{WMAP1} 
D.N. Spergel, L. Verde, H.V. Peiris, \textit{et al.}, \textit{Astrophys. J. Suppl. Ser.} {\bf 148} (2003) 175.

\bibitem{WMAP3} 
D.N. Spergel, R. Bean, O. Dor{\'e}, \textit{et al.}, \textit{Astrophys. J. Suppl. Ser.} {\bf 170} (2007) 377.

\bibitem{WMAP7} 
E. Komatsu, K.M. Smith, J. Dunkley, \textit{et al.}, \textit{Astrophys. J. Suppl. Ser.} {\bf 192} (2011) 18.

\bibitem{EisensteinBAO} 
D.J. Eisenstein and W. Hu, \textit{Astrophys. J.} {\bf 496} (1998) 605.

\bibitem{baodetection1} 
D.J. Eisenstein, I. Zehavi, D.W. Hogg, \textit{et al.}, \textit{Astrophys. J.} {\bf 633} (2005) 560.

\bibitem{baodetection2} 
S. Cole, W.J. Percival, J.A. Peacock, \textit{et al.}, \textit{Mon. Not. R. Astron. Soc.} {\bf 362} (2005) 505.

\bibitem{BethDR7} 
B.A. Reid, W.J.  Percival, D.J. Eisenstein, \textit{et al.}, \textit{Mon. Not. R. Astron. Soc.} {\bf 404} (2010) 60.

\bibitem{WillDR7} 
W.J. Percival, B.A.  Reid, D.J. Eisenstein,  \textit{et al.}, \textit{Mon. Not. R. Astron. Soc.} {\bf 401} (2010) 2148.

\bibitem{wiggle-z} 
C. Blake, E.A. Kazin, F. Beutler,  \textit{et al.}, \textit{Mon. Not. R. Astron. Soc.} {\bf 418} (2011) 1707.

\bibitem{Carbone/etal:2011} 
C. Carbone,  L. Verde, Y. Wang, and A. Cimatti, \textit{J. Cosmol. Astropart. Phys.} {\bf 2011}(03) (2011) 030.

\bibitem{JKGV10} 
R. Jimenez, T. Kitching, C. Pe{\~n}a-Garay and L. Verde, \textit{J. Cosmol. Astropart. Phys.} {\bf 2010}(05) (2010) 035.

\bibitem{dasi1} 
J.M. Kovac, E.M. Leitch, C. Pryke, \textit{et al.}, \textit{Nature} {\bf 420} (2002) 772.

\bibitem{dasi2} 
E.M. Leitch, J.M. Kovac, C. Pryke, \textit{et al.}, \textit{Nature} {\bf 420} (2002) 763.

\bibitem{ZaldarriagaHarari} 
M. Zaldarriaga and D.D. Harari, \textit{Phys. Rev.} {\bf D52} (1995) 3276.

\bibitem{SpergelZaldarriaga97} 
D.N. Spergel and M. Zaldarriaga, \textit{Phys. Rev. Lett.} {\bf 79} (1997) 2180.

\bibitem{Peiris/etal:2003} 
H.V. Peiris, E. Komatsu, L. Verde, \textit{et al.}, \textit{Astrophys. J. Suppl. Ser.} {\bf 148} (2003) 213.

\bibitem{CMBtaskforce} 
J. Bock, S. Church, M. Devlin, \textit{et al.}, arXiv:astro-ph/0604101 (2006).\aq{Please update publication details.}

\bibitem{BaumannWP} 
D. Baumann, M.G. Jackson, P. Adshead, \textit{et al.}, in \textit{CMB Polarization Workshop: Theory and Foregrounds: CMBPol Mission Concept Study}, AIP Conf. Ser., no.~1141, Eds.\ S. Dodelson \textit{et al.} (American Institute of Physics, New York, 2009), p.~10.

\bibitem{DETF} 
A. Albrecht, G. Bernstein, R. Cahn, \textit{et al.}, arXiv:astro-ph/0609591 (2006).\aq{Please update publication details.}

\bibitem{Peacockreport} 
J.A. Peacock, P. Schneider, G. Efstathiou, \textit{et al.}, Report by the ESA--ESO Working Group on Fundamental Cosmology, Eds.\ J.A.~Peacock \textit{et al.} (European Space Agency, 2006).

\end{thebibliography}
\end{document}